\documentclass[aps,prb,reprint,superscriptaddress]{revtex4-1}

\usepackage{graphicx}
\usepackage{amsmath,array,amssymb}
\usepackage{dcolumn}
\usepackage{bm}
\usepackage{hyperref}
\usepackage{color}

\DeclareMathOperator{\sech}{sech}

\graphicspath{{img/}}

\begin{document}

\title{Optical decoherence and spectral diffusion in an erbium-doped silica glass fiber featuring long-lived spin sublevels}

\author{Lucile Veissier}
\altaffiliation{Present address: Laboratoire Aim\'e Cotton, CNRS-UPR 3321, Univ. Paris-Sud, B\^at. 505, F-91405 Orsay Cedex, France}
\altaffiliation{The two first authors contributed equally to the work.}

\affiliation{Institute for Quantum Science and Technology, and Department of Physics \& Astronomy, University of Calgary, Calgary, AB, Canada}
\author{Mohsen Falamarzi}
\altaffiliation{The two first authors contributed equally to the work.}

\author{Thomas Lutz}

\author{Erhan Saglamyurek}
\altaffiliation{Present address: Department of Physics, University of Alberta, Edmonton, AB, Canada}
\affiliation{Institute for Quantum Science and Technology, and Department of Physics \& Astronomy, University of Calgary, Calgary, AB, Canada}

\author{Charles W. Thiel}
\affiliation{Department of Physics, Montana State University, Bozeman, MT, USA}

\author{Rufus L. Cone}
\affiliation{Department of Physics, Montana State University, Bozeman, MT, USA}

\author{Wolfgang Tittel}
\affiliation{Institute for Quantum Science and Technology, and Department of Physics \& Astronomy, University of Calgary, Calgary, AB, Canada}

\date{\today}

\begin{abstract}
Understanding decoherence in cryogenically-cooled rare-earth-ion doped glass fibers is of fundamental interest and a prerequisite for applications of these material in quantum information applications. Here we study the coherence properties in a weakly doped erbium silica glass fiber motivated by our recent observation of efficient and long-lived Zeeman sublevel storage in this material and by its potential for applications at telecommunication wavelengths. We analyze photon echo decays as well as the potential mechanisms of spectral diffusion that can be caused by coupling with dynamic disorder modes that are characteristic for glassy hosts, and by the magnetic dipole-dipole interactions between Er$^{3+}$ ions. We also investigate the effective linewidth as a function of magnetic field, temperature and time, and then present a model that describes these experimental observations. We highlight that the operating conditions (0.6 K and 0.05 T) at which we previously observed efficient spectral hole burning coincide with those for narrow linewidths (1 MHz) -- an important property for applications that has not been reported before for a rare-earth-ion doped glass.  
\end{abstract}


\maketitle

\section{Introduction}

Cryogenically-cooled rare-earth-ion (REI) doped materials offer unique spectroscopic properties, such as narrow optical linewidths and long-lived shelving levels that allow for spectral tailoring of their inhomogeneously broadened absorption lines. These properties are required simultaneously in order to implement many of the potential applications of REI-doped materials, including optical quantum memories \cite{tittel_photon-echo_2010, lvovsky_optical_2009}, as well as classical and quantum signal processing \cite{thiel_rare-earth-doped_2011,saglamyurek_integrated_2014}. Compared to crystals, the properties of REI's in amorphous hosts are generally very different because of the intrinsic disorder of the environment. This disorder comes with some advantages, in particular larger inhomogeneous broadening that is required for  high-bandwidth or spectrally multiplexed applications. Furthermore, the increased inhomogeneous broadening of electron and nuclear spin transitions can inhibit spin diffusion that leads to decoherence \cite{thiel2012optical}. This reduction was a key factor in our recent observation of efficient optical pumping into Zeeman sublevels with spin lifetimes reaching 30~s in an erbium-doped fiber \cite{saglamyurek_efficient_2015}. However, in addition to long-term storage mechanisms, another prerequisite for the above-mentioned applications is a narrow homogeneous linewidth. REI-doped amorphous materials generally exhibit much larger homogeneous linewidths at liquid helium temperatures compared to REI-doped crystalline hosts due to interactions with dynamic fluctuations in the environment that are traditionally modeled as bistable two-level systems (TLS) \cite{anderson_anomalous_1972,phillips_tunneling_1972}. 

Motivated by our recent observation of slow spin relaxation \cite{saglamyurek_efficient_2015} in a weakly-doped erbium-doped fiber that has allowed storing members of entangled photon pairs \cite{saglamyurek_quantum_2015}, we now investigate coherence properties of such a fiber with special attention to the regime of low magnetic fields where long Zeeman lifetimes were observed. More precisely, we report the magnetic field and temperature dependence of the optical coherence using two pulse photon echo (2PPE) and three pulse photon echo (3PPE) techniques. The fiber studied here has very similar composition and Er-doping concentration as the one used in our previous studies \cite{saglamyurek_efficient_2015,saglamyurek_quantum_2015}. In particular, both fibers feature long-lived persistent spectral holes with very similar characteristics.

Our paper is organized as follows. First, we discuss the results of the 2PPE measurements including the observed non-exponential echo decays in the context of spectral diffusion. We then present a detailed analysis describing the observed behaviors that combines elements from theoretical and semi-empirical models that have been developed  in the past for glassy and crystalline hosts. In agreement with results of previous studies \cite{macfarlane_optical_2006,staudt_investigations_2006}, we find that coupling with TLS significantly limits the coherence lifetimes to less than 1 $\mu$s even at high magnetic fields and temperatures as low as 600 mK.  
However, an important new finding is that the best coherence properties exist at weak magnetic fields of around 0.05~T, which correspond to the optimal field strength for persistent spectral hole burning in this material \cite{saglamyurek_efficient_2015}. This result is highly desirable for applications requiring both long coherence times as well as long spin state lifetimes.

\section{Experimental details}

Our silica glass fiber (INO Canada, S/N 404-28565) is 25 m long and contains erbium, aluminium, germanium and phosphorus co-dopants. The Er and Al doping concentrations in the core are 80 ppm and 1800 ppm, respectively, and the concentrations of G and P are unknown. The fiber is cooled to  temperatures as low as $T=600$~mK using an adiabatic demagnetization refrigerator (please note that, unlike fluoride and tellurite glass fibers, silica glass fibers are easy to use at cryogenic temperatures), resulting in an optical depth of $\alpha L = 1.6$ at  $\lambda=1536$~nm wavelength. Magnetic fields $B$ of up to 2~T are applied by means of a superconducting solenoid.

To perform the 2PPE (or 3PPE) measurements, we used two fiber-coupled electro-optic modulators to generate excitation pulses from a continuous wave external-cavity diode laser operating at $\lambda=1536$~nm. The light was then amplified using an erbium-doped fiber amplifier (EDFA). In order to suppress spontaneously emitted and amplified light from the EDFA, we filtered the output in polarization with a polarizing beam-splitter, in frequency with a 1~nm bandpass filter, and in time using an additional intensity modulator before sending the excitation pulses into the Er$^{3+}$-doped fiber. The first pulse was 4~ns long, and the duration of the second (and third in case of 3PPE) pulse was 8~ns. All pulses emerging from the fiber were detected with an amplified photodetector (Newport AD-200xr). Since any persistent hole-burning will affect the echo signal strength, we ensured that the repetition period of the experiment exceeds the characteristic (magnetic field dependent \cite{saglamyurek_efficient_2015}) persistence lifetime.

\section{Two-pulse photon echo measurements -- analysis and model}

\begin{figure}[t]
\centering
\includegraphics[width=0.9\columnwidth]{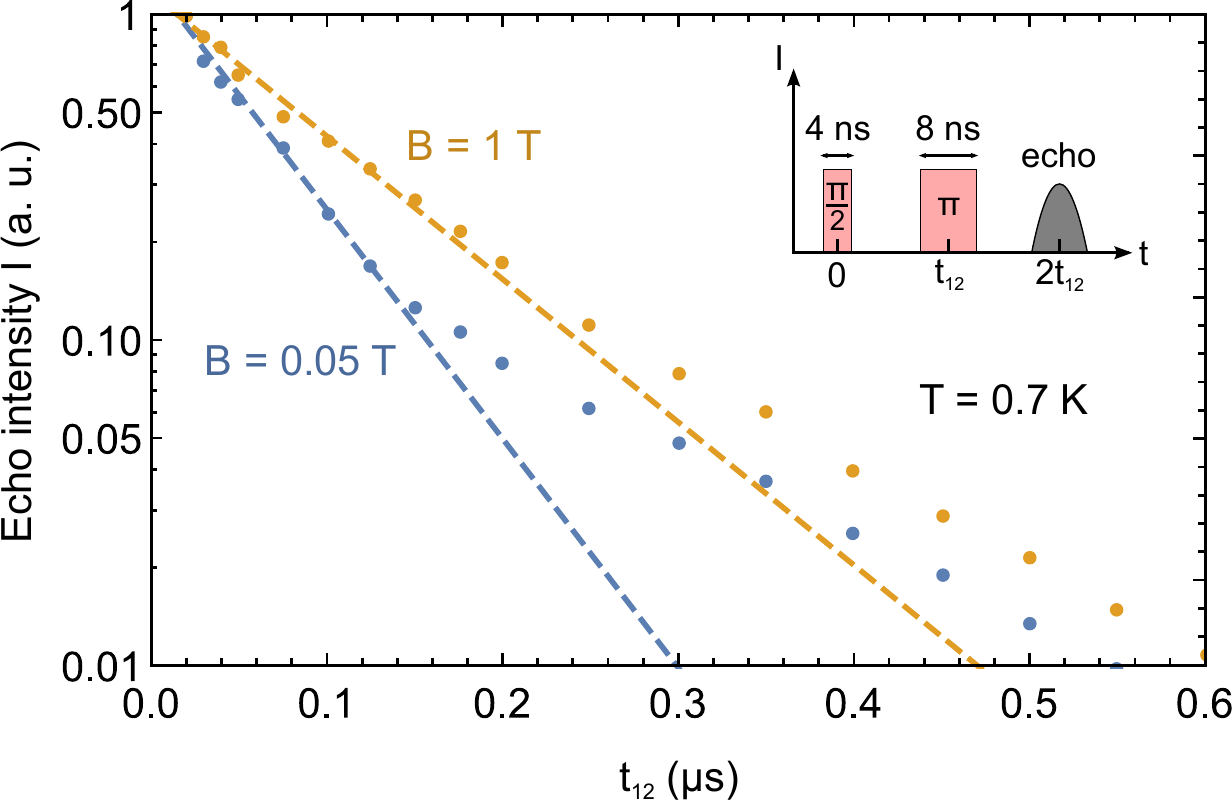}
\caption{Two-pulse photon echo peak intensity as a function of the time delay between the two excitation pulses at $T=0.7$~K for $B= 0.05$~T (blue dots) and $B=1$~T (yellow dots). The pulse sequence is shown in the inset. The experimental data are fitted by a single-exponential function (dashed lines) with characteristic decay constants $T_2 = 247 \pm 14$ ns and $T_2=396 \pm 16$ ns.}
\label{fig:echo_decay}
\end{figure}

The optical coherence properties of the fiber can be extracted from two-pulse photon echo measurements. In a 2PPE sequence, two short pulses, separated by a waiting time $t_{12}$ are sent into an inhomogeneously-broadened ensemble of absorbers. This gives rise to the emission of an echo at time $t_{12}$ after the second pulse. The variation of its intensity as a function of $t_{12}$  
\begin{equation}
I(t_{12}) = I_0 \, e^{-4 \pi \Gamma_{\rm h} t_{12}}
\end{equation}
reveals the homogeneous linewidth $\Gamma_{\rm h}$ (which is inversely proportional to the coherence lifetime: $\Gamma_{\rm h}=(\pi T_2)^{-1}$) assuming that all absorbers have the same coherence properties. Fig.~\ref{fig:echo_decay} shows typical examples of the echo intensity as a function of the pulse separation $t_{12}$ for $B= 0.05$~T and $B=1$~T. The echo decays clearly show a non-exponential behavior: they deviate from a single exponential fit (see dashed lines) roughly at $t_{12} \geq 0.2 \; \mu$s, and feature a slower decay after this point. This differs from what has been observed in previous investigations of coherence properties in Er-doped fibers, where simple exponential decays were reported \cite{macfarlane_optical_2007,staudt_investigations_2006} -- most likely because the dynamic range of the echo decays in theses past studies was smaller.
In our case, the experimental echo decays can be fitted by the sum of multiple exponential functions, which could suggest the presence of several classes of ions with distinct coupling strengths to their environment, or of several distinct perturbing processes. However, these explanations are not likely for an amorphous host where a random distribution of different site environments is usually assumed.

\begin{figure}[t]
\centering
\includegraphics[width=1\columnwidth]{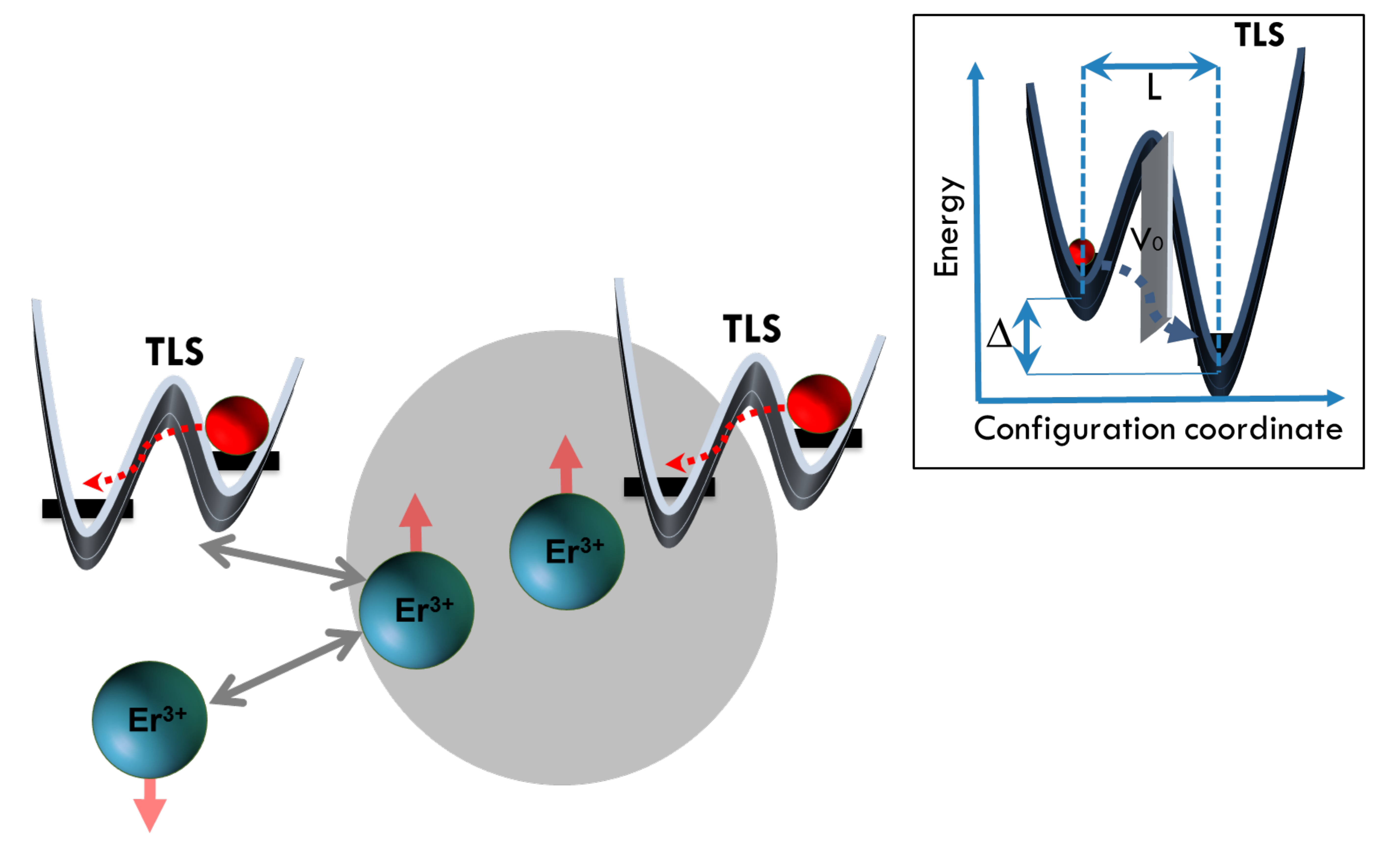}
\caption{Spectral diffusion processes in a rare-earth ion doped glass. The local environments of the optically probed Er$^{3+}$ ions in the center of the grey circle are perturbed by either direct interaction with fluctuations in the state of neighboring TLS (depicted on the right-hand side), or interaction with fluctuations in the state of other Er$^{3+}$ spins in the local environment driven by TLS or another Er$^{3+}$ ion. Inset: double-well potential-energy structure of the TLS, with asymmetry energy $\Delta$, barrier energy $V_0$ and well separation $L$.}
\label{fig:SDscheme}
\end{figure}

Due to the amorphous nature of the fiber, we expect a broad, continuous distribution of coherence-limiting processes to affect the Er$^{3+}$ ions. Of particular importance are changes in the local electric or magnetic field at the Er$^{3+}$ ion position; they result in time-dependent shifts of its optical transition frequency -- so-called spectral diffusion -- which is taken into account by using an effective homogeneous linewidth $\Gamma_{\rm eff}$. As depicted in Fig.~\ref{fig:SDscheme}, these shifts can arise due to interactions with neighboring TLS's (or tunneling modes) and Er$^{3+}$ spins, each described by two states featuring a certain energy splitting and random population fluctuations at a certain local rate \cite{broer_low-temperature_1986,zyskind_determination_1990}. 

Two-level systems\cite{anderson_anomalous_1972,phillips_tunneling_1972} are intrinsic to amorphous materials  and are present with a continuous distribution of flip rates $R$ and energy splittings $E$ that both depend on the asymmetry energy $\Delta$, well separation $L$, and barrier energy $V_0$ (see inset of Fig.~\ref{fig:SDscheme}), as well as on the magnetic field in certain cases (so-called spin-elastic or magnetic TLS's) \cite{sun_exceptionally_2006,staudt_investigations_2006,macfarlane_optical_2007}. 
Thus, the calculation of the intensity of a 2PPE must include integration over a continuous distribution of rates and energy splittings:
\begin{equation}
I(t_{12}) = I_0 \int \int e^{-4 \pi \Gamma_{\rm eff} (R,E,T,t) t_{12}} dR \, dE \; ,
\label{eq:2PEintensity}
\end{equation}
where
\begin{equation}
\Gamma_{\rm eff} (R,E,T,t) = \Gamma_0(T) +\Gamma_{\rm SD}(E,T) P(R,E) (1-e^{-Rt})\; .       
\end{equation}
$\Gamma_0(T)$ is the intrinsic homogeneous linewidth (without spectral diffusion), and the spectral diffusion linewidth $\Gamma_{\rm SD}$ is given by
\begin{equation}
\Gamma_{\rm SD}(E,T) = \Gamma_{\rm max} \sech ^2 \left( \frac{E} {2 k T} \right)  \, ,
\label{eq:GammaSD}
\end{equation}
where $\Gamma_{\rm max}$ is the maximum possible linewidth broadening caused by spectral diffusion, $k$ the Boltzmann constant, and $T$ the temperature \citep{black_spectral_1977}. Furthermore, the probability distribution for finding a TLS with energy $E$ and flipping rate $R$, $P(R,E)$, is given by
\begin{equation}
P(R,E) = \frac{1}{R \sqrt{1 - R/R_{\rm max}(E)}} \; ,
\end{equation}
with $ R_{\rm max}(E) \propto E^3 \coth \left( \frac{E}{2 k T} \right)$\cite{jackle_ultrasonic_1972}. In the case of interactions with a magnetic TLS or the spin of another Er$^{3+}$ ion, the energy $E$ is given by $g \mu_B B$, where $g$ is the effective $g$-value for the perturbing TLS or Er$^{3+}$ ion, $ \mu_B$ the Bohr magneton, and $B$ the applied magnetic field.
Using Eq.~\ref{eq:2PEintensity}, one can fit $\Gamma_{\rm eff}$ to the individual echo decays by summing over five individual processes: spectral diffusion due to direct interaction of the probed Er$^{3+}$ ions with non-magnetic (process 1) or magnetic (process 2) TLSs; spectral diffusion due to interaction of the probed Er$^{3+}$ ions with pairs of coupled Er$^{3+}$ spins in the environment that randomly exchange spin states (Er$^{3+}$-Er$^{3+}$ spin flip flops) (process 3); and spectral diffusion due to interaction of the probed Er$^{3+}$ ions with Er$^{3+}$ spins that are strongly coupled to magnetic (process 4) or non-magnetic (process 5) TLSs that drive spin flips (i.e. Er$^{3+}$-TLS flip flops).
Due to the data set being limited in size and quality as well as computational complexity, we did not succeed in fitting all echo decays with one unique set of model parameters. However, as we will describe next, restricting the coherence-limiting processes to (1), (3) and (4), we were able to reproduce the magnetic field and temperature dependence of the single-exponential fits to our data shown in Fig.\ref{fig:echo_decay} for the effective homogeneous linewidth $\Gamma_{\rm eff}(B,T)$.

\section{Temperature and magnetic field dependence of the effective homogeneous linewidth}

To extract the effective homogeneous linewidth as a function of magnetic field and temperature, we fit all measured echo decays using the single exponential function described in Eq.~\ref{eq:2PEintensity} (after replacing $\Gamma_{\rm h}$ with $\Gamma_{\rm eff}$). These fits describe the coherence data over the first decade of the decays, thereby restricting the assessment of coherence-limiting processes to that region of the data; nevertheless, we should note that the first decade of the decay represents the dominant decoherence mechanisms that are primarily responsible for the performance in applications, with the different behaviors observed at longer times resulting from either higher-order processes and correlations or groups of minority ion sites in the material.

\begin{figure}[t]
\includegraphics[width=0.9\columnwidth]{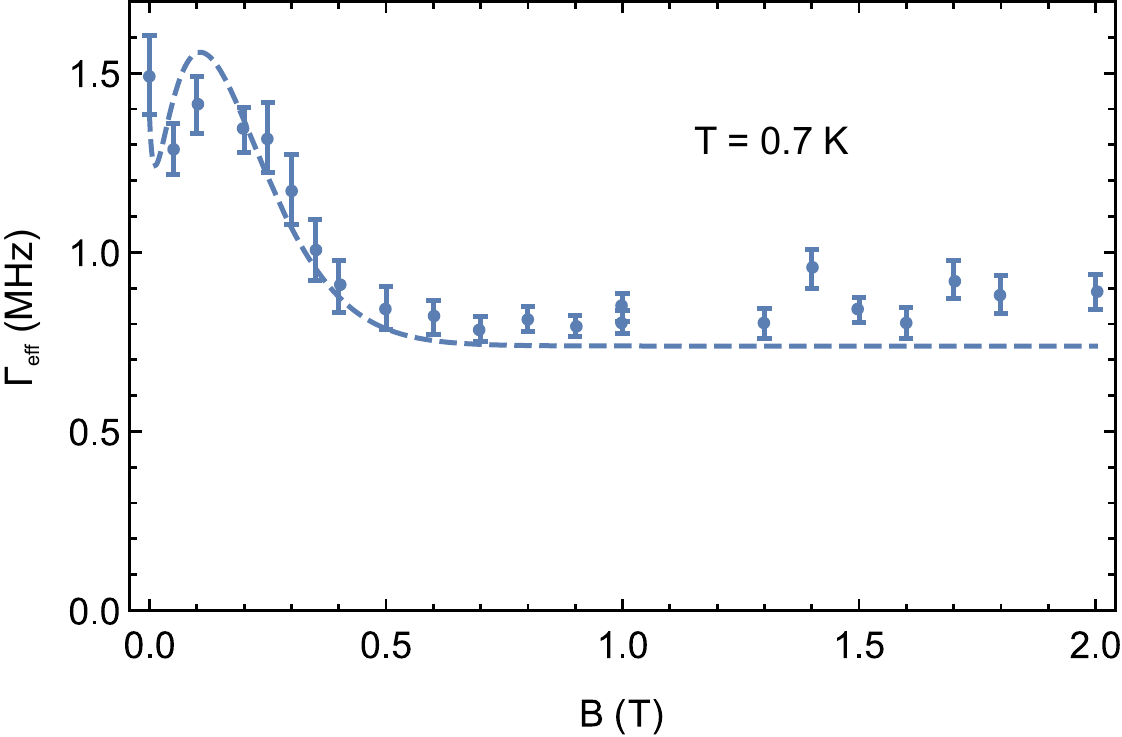}
\caption{Magnetic field dependence of the effective homogeneous linewidth $\Gamma_{\rm eff}$ at $T=0.7 \pm 0.05$~K. The dashed lines shows the theoretical prediction of Eq.~\ref{eq:model} with the set of parameters in Table~\ref{table:parameters}. }
\label{fig:b-dep}
\end{figure}

\begin{figure}[t]
\centering
\includegraphics[width=.9\columnwidth]{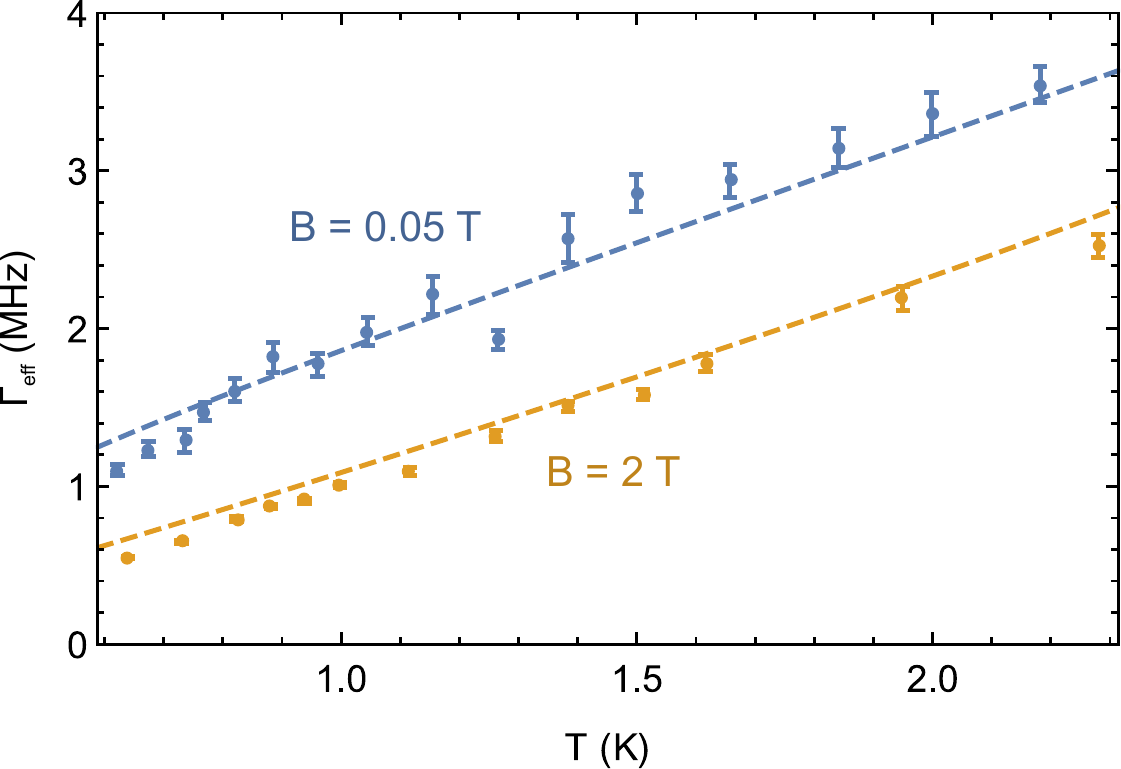}
\caption{Temperature dependence of the effective homogeneous linewidth $\Gamma_{\rm eff}$ at $B=0.05$~T and $B=2$~T. The dashed lines shows the theoretical prediction of Eq.~\ref{eq:model} with the set of parameters in Table~\ref{table:parameters}.}
\label{fig:t-dep}
\end{figure}

In Figs.~\ref{fig:b-dep} and \ref{fig:t-dep} we plot examples of the experimentally obtained effective linewidth $\Gamma_{\rm eff}$ as a function of magnetic field for a fixed temperature and as a function of temperature for two different magnetic fields. In the case of varying magnetic field (Fig.~\ref{fig:b-dep}), we observe two components. The first one is magnetic field independent and has been attributed to dephasing due to non-magnetic, or elastic, TLS \cite{macfarlane_optical_2007,staudt_investigations_2006}. At 0.7~K its contribution to the Er$^{3+}$ homogeneous linewidth in our fiber is roughly 0.75 MHz. The second component is magnetic field dependent: it dominates at small magnetic fields and is rapidly suppressed as the field increases. This behavior has been attributed to spectral diffusion due to the interaction of Er$^{3+}$ ions with magnetic TLS \cite{macfarlane_optical_2007,staudt_investigations_2006} 
, or other Er$^{3+}$ ions in the environment \cite{sun_exceptionally_2006}. The latter process is regularly observed in Er$^{3+}$-doped crystals \cite{bottger_optical_2006,thiel_optical_2010,bottger_decoherence_2016}. In the region of low magnetic field, we observe a local minimum at $B\approx 0.05$~T, followed by a local maximum at $B\approx 0.15$~T. This behavior is similar to what has been observed in Er$^{3+}$-doped crystals, such as Er$^{3+}$:LiNbO$_3$ and Er$^{3+}$:KTP \cite{thiel_optical_2010,bottger_decoherence_2016}, and can be explained by the competition between two mechanisms: the interaction between Er$^{3+}$ ions, which decreases with $B$, and the interaction between Er$^{3+}$ ions and resonant TLS, which increases with $B$ according to the TLS density of states \cite{saglamyurek_efficient_2015}. Note that the overall effect of the spectral diffusion caused by Er-Er interactions i.e. processes (3,4,5) is 0.7 MHz at 0.1 T and 0.7 K in our 80 ppm Er-doped fiber, whereas it is about 40 kHz at the same magnetic field and at 3 K in a 80 ppm Er$^{3+}$:LiNbO$_3$ crystal\cite{thiel_optical_2010}. Thus, this contribution is larger in the fiber, but due to the broadening of the spin transition reducing spin flip flops, it is suppressed more rapidly with the application of a magnetic field.

Investigating $\Gamma_{\rm eff}$  at low and high magnetic fields as a function of temperature (Fig.~\ref{fig:t-dep}), we observe that both components increase nearly linearly with temperature. This agrees with the characteristic behavior of TLS-limited coherence \cite{Phillips1987}, in which the effective linewidth is proportional to $ T^n$, with values of $n$ reported between 1 and 1.5 \cite{macfarlane_optical_2007,broer_low-temperature_1986,staudt_investigations_2006}. 

\subsection{Spectral diffusion model}

Next, we fit the effective linewidths using a model that takes into account the three processes described previously (1, 3, and 4). 
Our model combines what has been proposed for amorphous media \cite{Phillips1987} and crystals (in particular in the case of Er:YSO \cite{bottger_optical_2006}). We write the coherence lifetime as
\begin{equation}
\begin{aligned}
T_2 = & \frac{2 \,  (\Gamma_0+\alpha_0 T^n)}{\Gamma_{\rm SD} R} \left(\sqrt{1 + \frac{\Gamma_{\rm SD} R}{\pi (\Gamma_0+\alpha_0 T^n)^2}}-1\right).
\end{aligned}
\label{eq:model}
\end{equation}
Here, $\Gamma_0$ is the homogeneous linewidth at $T=0$~K, $\alpha_0$ is a constant describing the direct interaction with non-magnetic TLS (process 1), and $1 \leq n \leq 1.5$. The term  $\frac{1}{2}\Gamma_{\rm SD} R$ represents the shortening of the coherence lifetime through spectral diffusion caused by magnetic dipole-dipole interactions (since the Zeeman energy level splittings are on the order of kT, only spins are thermally activated and thus, the dynamics are dominated by magnetic dipole-dipole interactions between co-dopants) with surrounding Er$^{3+}$ spins, which are flipped by neighboring Er-ions or TLS (processes 3 and 4). The associated flipping rate $R$ can be described by
\begin{equation}
\begin{aligned}
R(B,T) = & \frac{ \alpha_{1} }{\Gamma_{\rm S}^0 +  \gamma_{\rm S} B} \sech ^2 \left( \frac{g_{\rm env} \mu_{B} B} {2 k T} \right) \\
& +  \alpha_{2} B T  \, .
\end{aligned}
\label{eq:rate}
\end{equation}
The first term corresponds to spin flip-flops between Er$^{3+}$ ions, with a coupling coefficient $\alpha_1$ (process 3). We include broadening of the inhomogeneous linewidth $\Gamma_{\rm S}$ of the spin transition with magnetic field, i.e. $\Gamma_{S_0}=\Gamma_{\rm S}^0+\gamma_{\rm S} B$, with $\Gamma_{\rm S}^0=1.5$ GHz and $\gamma_{\rm S} = 150$ GHz/T, as observed in our previous work on spin relaxation in erbium-doped fiber \cite{saglamyurek_efficient_2015}. The second term corresponds to flip flops with magnetic TLS, with a coupling coefficient $\alpha_2$ (process 4). While this effect is in some ways analogous to the coupling with phonons in crystals, we observed in the same previous work that it is proportional to $B$, i.e. has a different dependence on the magnetic field due to the difference in the density of states of the TLS \cite{buchenau_low-frequency_1986}. The flip-flop rate increases with magnetic field for $B>0.05$~T, which opposes the reduction of the linewidth $\Gamma_{\rm SD}$ with $B$, hence giving rise to the local maximum observed in the magnetic field dependence of the homogeneous linewidth (see Fig. \ref{fig:b-dep}). Fitting Eq.~\ref{eq:model} to the magnetic field and temperature dependence of the homogeneous linewidth $\Gamma_{\rm eff}(B,T)$ results in the parameters listed in Table~\ref{table:parameters}. The values of $\Gamma_{\rm S}^0$ and $\gamma_{\rm S}$ have been fixed to the values found in our previous investigation of 1.5 GHz and 150 GHz/T, respectively \cite{saglamyurek_efficient_2015}.

\begin{table}[t]
\begin{center}
\begin{tabular}{|c|c|}
\hline
$\Gamma_0$ & $ (0.0 \pm 0.5)$ MHz \\
$\alpha_0$ & $ (1.1 \pm 0.5)$ MHz/T$^n$ \\
$n$ & $1.1 \pm 0.4$ \\
$g_{\rm env}$ & $ 14.4 \pm 1.6$  \\
$\alpha_1 /\Gamma_{\rm max}$  &  $11  \pm 5$ GHz \\
$\alpha_2 / \Gamma_{\rm max}$ &  $348 \pm42$ (T K)$^{-1}$  \\
\hline
\end{tabular}
\end{center}
\caption{Parameters resulting from the 2-dimensional fit of Eq.~\ref{eq:model} to $\Gamma_{\rm eff}(B,T)$.}
\label{table:parameters}
\end{table}
 
The agreement between our model and the experimental data is exemplified by the data sets shown in Figs.~\ref{fig:b-dep} and \ref{fig:t-dep}. Furthermore, the value of 1.1 for the exponent $n$ is in agreement with previous work \cite{macfarlane_optical_2007,broer_low-temperature_1986,staudt_investigations_2006}, and the value of $g_{\rm env}=14.4$, while large, is within the allowed range from 0 to 18 for the $^4$I$_{15/2}$ levels of Er$^{3+}$ ions \cite{asatryan2007electron,ammerlaan2001zeeman,nolte1997epr,bravo1993electron,reynolds1972epr,belyaeva1968anisotropy}. 
We also note that the rate of Er$^{3+}$ spin flips, characterized by $\alpha_1$ and $\alpha_2$, is on the order of MHz to GHz.However, we reported spin relaxation (i.e. spin flip) rates on the order of Hz in our previous study of persistent spectral hole burning in a similar fiber\cite{saglamyurek_efficient_2015}. We believe that this large difference in behavior can be explained by the broad distribution of spin relaxation rates spanning the entire range from Hz to GHz. The coherence lifetime is limited by spectral diffusion that occurs at the fastest rate whereas the possibility for persistent spectral hole burning is determined by Er$^{3+}$ ions with the slowest rates; in addition to causing spectral diffusion, the small fraction of rapidly relaxing ions leads to the observed temperature-dependent limit on the maximum hole depth.

Overall, the coherence properties improve at low temperature and high magnetic fields, with an optimum of $\Gamma_{\rm eff}=0.55$~MHz at $T=0.64$~K and $B =2$~T (see Fig.~\ref{fig:t-dep}). However, the broadening of the effective linewidth at magnetic fields of around 0.05 T (at which persistent spectral hole burning is possible) is small enough to allow linewidths of approximately 1 MHz at $T=0.64$~K (see Fig.~\ref{fig:t-dep}). This property has not been previously observed \cite{macfarlane_optical_2006,staudt_investigations_2006}, possibly due to differences in co-dopants (phosphorus in our fiber and lanthanum in the fiber studied, e.g., in \cite{macfarlane_optical_2006}). Another interesting observation is that the value for $\Gamma_0$ predicts a very narrow linewidth for $T\rightarrow 0$~K. However, note that this is only an extrapolation of our observations down to $T=0.6$~K, and it is likely that other processes limit the linewidth to a constant value at lower temperatures.

\section{Three-pulse photon echo measurements -- spectral diffusion at large timescales}

Finally, to investigate spectral diffusion at timescales comparable to the excited-state lifetime of Er$^{3+}$ (11 ms), 3PPE measurements were carried out at a temperature of $T=0.76$~K. We chose two values of the magnetic field: $B=0.06$~T for which the coherence lifetime exhibits a local maximum, and $B=2$~T for which the effect of spectral diffusion due to Er$^{3+}$-Er$^{3+}$ interaction is small and the coherence is limited by the interaction with TLS. For these measurements, the separation time $t_{12}$  between the first two pulses was held constant at 50 ns and the echo intensity was measured as a function of the time $t_{23}$ between the second and third pulse, with $t_{23}$ varying between 0.001 ms and 35 ms. The echo intensity is given by
\begin{equation}
\begin{aligned}
 I & \left(t_{12},t_{23} \right)= \\
 & I_0 \left\lbrace \; e^{- t_{23}/T_1}  + \frac{\beta}{2} \,\frac{ T_Z}{T_Z - T_1 } \left( e^{ - t_{23}/T_Z} -  e^{ - t_{23}/T_1} \right) \right\rbrace ^2 \\
& \times e^{ -4 \, t_{12}\pi \Gamma_{\rm eff}(t_{12},t_{23}) }.
\end{aligned}
\label{eq:3PPE_intensity}
\end{equation} 
where $I_0$ is a scaling coefficient and $\beta$ the branching ratio from the excited-level (with lifetime $T_1$) to the other Zeeman sublevel of the ground state (with lifetime $T_Z$) \cite{bottger_optical_2006}. A fit to our data with $T_1=11$~ms yields the effective homogeneous linewidth $\Gamma_{\rm eff}$ that depends on $t_{12}$ and $t_{23}$.

\begin{figure}[t]
\centering
\includegraphics[width=.9\columnwidth]{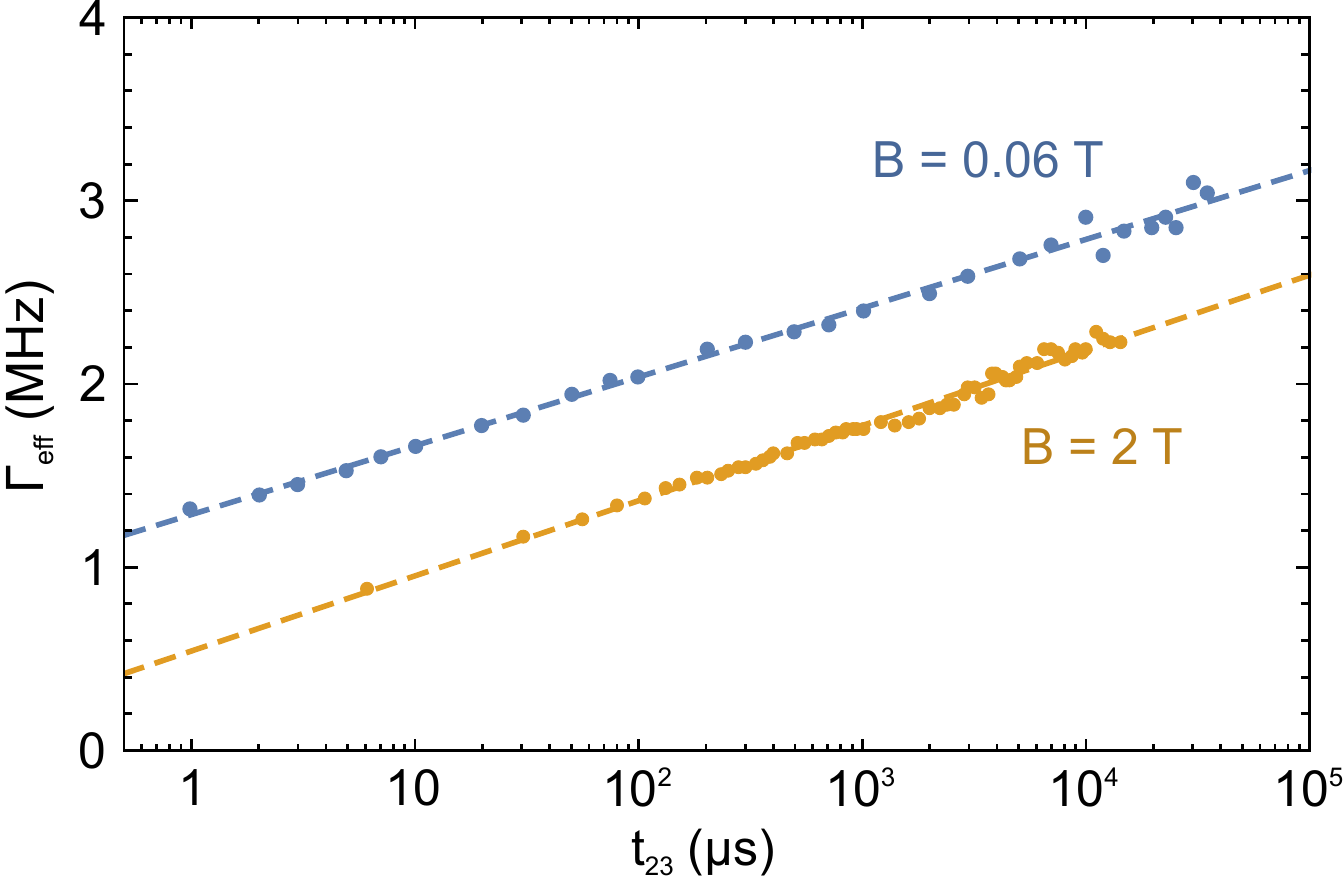}
\caption{Effective homogeneous linewidth as a function of $t_{23}$, derived from 3PPE measurements at $T=0.76$~K at $B=0.06$~T (empty circles) and  $B=2$~T (filled circles). The experimental data is fitted by Eq.~\ref{eq:3PPE_intensity} with $\gamma =0.376$~MHz/decade  and  0.410 MHz/decade at $B = 0.06$~T and 2 T respectively.}
\label{fig:3ppe}
\end{figure}

The measurement results are shown in Fig.~\ref{fig:3ppe}, which also includes fits of the expected logarithmic dependence of $\Gamma_{\rm eff}$ on $t_{23}$ due to coupling to TLS in fiber \cite{black_spectral_1977,silbey_time_1996}:
\begin{equation}
\Gamma_{\rm eff}(t_{12},t_{23})= \Gamma(t_0)+\gamma \:\log_{10} \left( \frac{t_{23}}{t_{0}}\right).
\label{eq:SD_TLS}
\end{equation}
Here, $\Gamma(t_0)$ is the effective linewidth at the minimum value of $t_{12}+t_{23} \equiv t_{0}$, and $\gamma$ is a coupling coefficient. The value of $\Gamma(t_0)$ is set to the one we measured in the 2PPE at the same magnetic field. We find excellent agreement for $\gamma =0.376$~MHz/decade  at $B = 0.06$~T and $\gamma = 0.410$~MHz/decade at 2 T. We find that, as in the case of short waiting times, our Er$^{3+}$-doped fiber features better coherence properties for small magnetic fields. Furthermore, effective linewidths at long waiting times barely improve when increasing the field to 2~T. This implies that spectral tailoring of the absorption profile with MHz resolution is possible even after large delays of up to hundreds of ms. In particular, this allows our fiber to be used as a quantum memory for light \cite{saglamyurek_quantum_2015}.

\section{Conclusion}

In conclusion, we have experimentally and theoretically investigated decoherence and spectral diffusion over a large range of timescales in an erbium-doped fiber similar to the one in which persistent hole burning has recently been demonstrated \cite{saglamyurek_efficient_2015}. Our model combines the semi-empirical model framework developed for amorphous hosts, where spectral diffusion is caused by the interaction with two-level systems, and the theoretical framework successfully applied previously for Er$^{3+}$-doped crystals, where spectral diffusion is due to Er$^{3+}$-Er$^{3+}$ magnetic dipole interactions. Most importantly, we found coherence lifetimes at small magnetic fields, at which Zeeman lifetimes can be as long as seconds, that are comparable to those at high magnetic fields. This is crucial for applications in the field of quantum information processing, in particular quantum memory for light \cite{saglamyurek_quantum_2015}.

\section{Acknowledgments}

The authors thank Daniel Oblak and Neil Sinclair for discussions and acknowledge support from Alberta Innovates Technology Futures (ATIF), the National Engineering and Research Council of Canada (NSERC), the US National Science Foundation (NSF) under award nos. PHY-1415628 and CHE-1416454, and the Montana Research and Economic Development Initiative. Furthermore, W.T. acknowledges support as a Senior Fellow of the Canadian Institute for Advanced Research (CIFAR).

%

\end{document}